\pdfoutput=1
\documentclass[fleqn]{article}
\usepackage{amsmath,amssymb}
\usepackage{epsfig,graphicx}

\catcode`@=11 \@addtoreset{equation}{section} \catcode`@=12

\begin{document}
\title{\vskip-1.7cm \bf  On suppression of topological transitions in quantum gravity}
\date{}
\author{A.O.Barvinsky}
\maketitle
\hspace{-8mm} {\,\,\em Theory Department, Lebedev
Physics Institute, Leninsky Prospect 53, Moscow 119991, Russia}

\begin{abstract}
We discuss the effect of dynamical suppression for a special class of topological configurations in cosmology, which occur in Euclidean quantum gravity (EQG) when the latter is viewed as the derivative of the physical theory in the Lorentzian signature spacetime. At the topological level EQG inherits from the Lorentzian theory the arrow of time and incorporates special junction conditions on quantum fields whose quantum fluctuations make the contribution of such topologies vanishing. This effect is more general than the recently suggested conformal mechanism of suppression of vacuum no-boundary instantons in the microcanonical statistical sum of quantum cosmology driven by a conformal field theory (CFT). In contrast to conformal properties of the CFT driven cosmology, this effect is based only on short-distance behavior of local boson fields and Pauli principle for fermions. Application of this effect in the CFT cosmology treated as initial conditions for inflationary Universe suggests the thermal nature of the primordial power spectrum of the CMB anisotropy. This can be responsible for a thermal contribution to the red tilt of this spectrum, additional to its conventional vacuum component.
\end{abstract}

\maketitle
\section{Introduction}
It is well known that summation over spacetime topologies is an important part of Euclidean quantum gravity (EQG) -- the concept underlying the theory of initial conditions in quantum cosmology \cite{noboundary} and physics of baby universes \cite{babyu1,babyu2}. The latter in its turn used to underlie the old mechanism of vanishing cosmological constant and big fix of fundamental constants of nature \cite{bigfix}. EQG interpreted as a fundamental theory a priori admits this summation and the corresponding topological transitions. On the other hand, in the Lorentzian signature spacetime temporal evolution with changing topology of spatial slices was demonstrated to be inconsistent in view of back reaction of infinitely intensive flashes of matter radiation emanating from the relevant spacetime bifurcation points \cite{AndersonDeWitt}.

In contrast to Lorentzian theory EQG is more flexible to accommodate various topologies, and except certain geometrical and kinematical restrictions (see for example \cite{kinematic}) it does not seem to contain dynamical mechanisms suppressing topological transitions. Nevertheless, there exists a counterexample to this statement -- it was recently demonstrated \cite{slih,why} that the EQG path integral for the microcanonical statistical sum in cosmology can suppress the contribution of instantons with the topology of the Hartle-Hawking no-boundary configurations \cite{noboundary}. This happens for the cosmological model driven by conformal field theory (CFT) -- quantum matter conformally coupled to gravity.

This model is interesting for several reasons. First, in view of suppression of the above type it eliminates a very anti-intuitive situation when infinitely large universes with a vanishing effective Hubble factor $H=0$ are infinitely more probable than the universes with a finite $H$ \cite{slih}. This is the situation with the Hartle-Hawking instantons having a {\em negative} action inverse proportional to $H^2$. Second, when properly modified (with the primordial cosmological constant $\Lambda=3H^2$ replaced by a composite field decaying in the slow roll regime) this model can generate inflation \cite{slih,bigboost}. And finally, as was noticed in \cite{DGP/CFT} the CFT driven cosmology provides perhaps the first example of the initial quantum state of the inflationary Universe, which has a thermal nature of the primordial power spectrum of cosmological perturbations. This suggests a new mechanism for the red tilt of the CMB anisotropy, complementary to the conventional mechanism which is based on a small deviation of the inflationary expansion from the exact de Sitter evolution \cite{ChibisovMukhanov}. Simple estimates show that the thermal imprint of this initial state can constitute a considerable or even dominant part of this red tilt. It is currently getting measured by Planck with ever growing precision at the pivotal wavelength scale $\sim$ 500 Mpc. Therefore, potentially this CFT model can be applied for experimental verification of the relative vacuum and thermal components of the red CMB spectrum at subhorizon scales.

In this paper we show that the suppression mechanism for topological configurations of the Hartle-Hawking type, which was found in the CFT driven cosmology, is much more general and, in fact, applies to all quantum fields independently of their conformal properties. Though this mechanism is realized within the EQG formalism, its origin can be traced back to the physical setting in the Lorentzian spacetime. EQG formalism derived from the Lorentzian theory inherits the arrow of time which, in its turn, enforces certain junction conditions for quantum fields at the points of bifurcating topology. These junction conditions infinitely suppress the contribution of such topologies within the one-loop approximation for the EQG path integral. For local boson fields this suppression follows from their ultraviolet (UV) behavior, while for fermions it turns out to be a direct consequence of the Pauli principle.

\begin{figure}[h]
\centerline{\includegraphics[width=8cm]{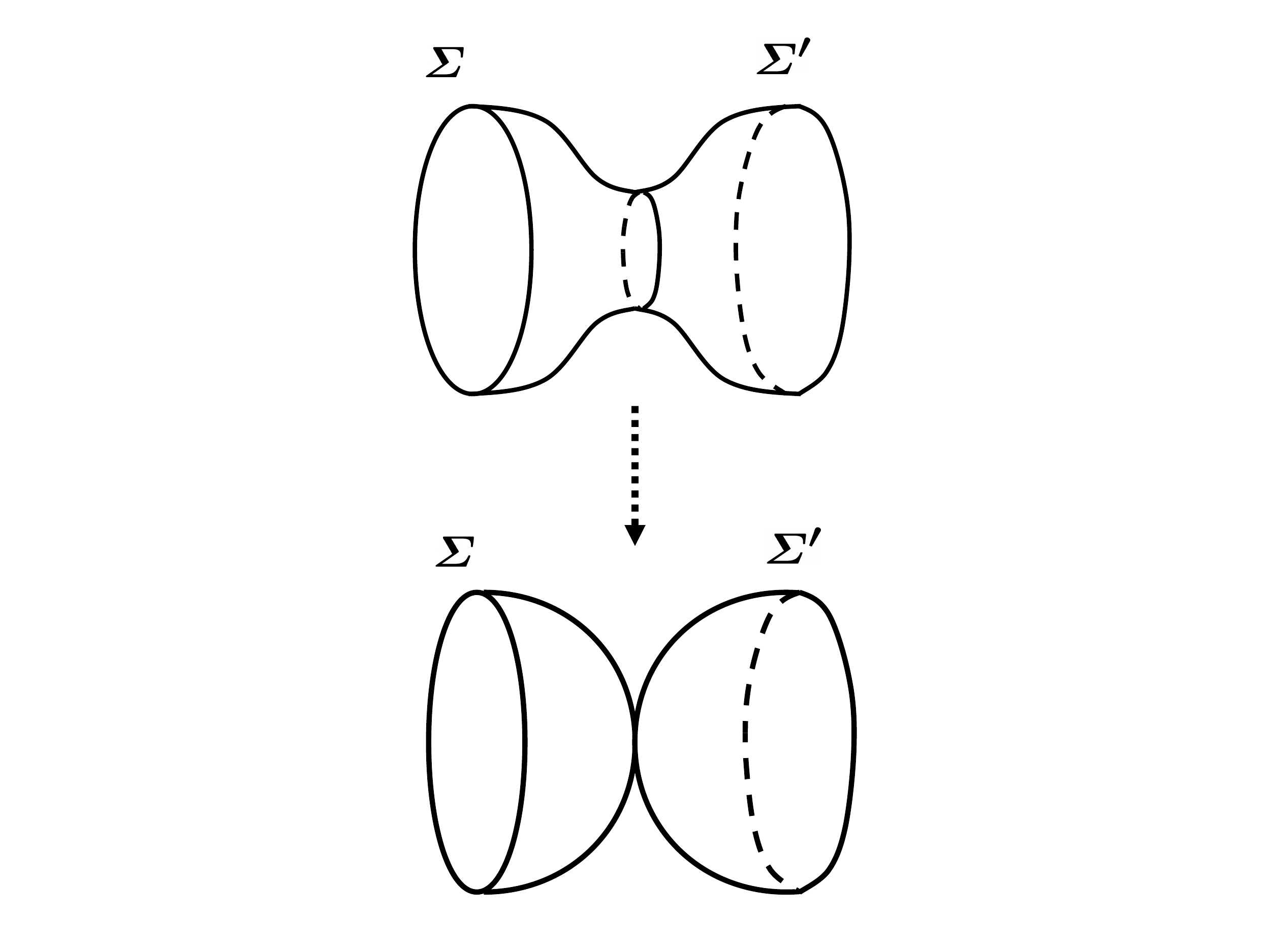}} \caption{\small
 \label{Fig.3}Pinching procedure for a segment of $S^3\times S^1$ spacetime between two closed hypersurfaces $\Sigma$ and $\Sigma'$ with the $S^3$ topology. The intermediate $S^3$ section contracts to a point which becomes a regular internal point of each of the two smooth disconnected parts.}
\end{figure}

\begin{figure}[h]
\centerline{\includegraphics[width=9cm]{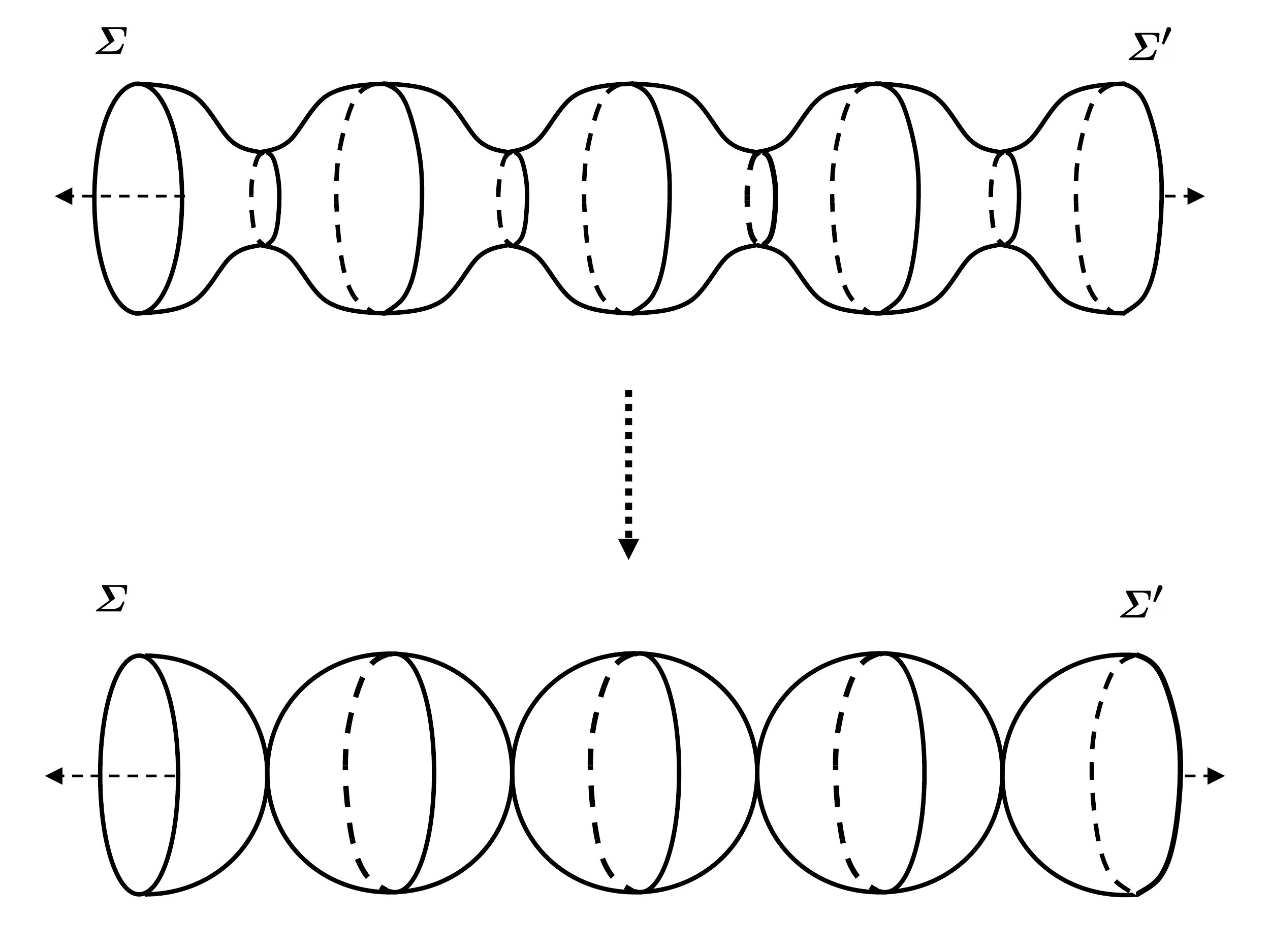}} \caption{\small
\label{Fig.3a} The origin of the multiple sphere topology for a segment of the periodic instanton $S^3\times S^1$. At the upper part of the figure the double-sided arrow (or axis) of time indicates the direction of periodic identification on the circle $S^1$ of $S^3\times S^1$. At the bottom part this arrow passes through pinching points.}
\end{figure}

The organization of the paper is as follows. In Sect.2 we recapitulate the suggestions of the work \cite{why} -- using as a first principle canonical quantization of gravity in the Lorentzian signature spacetime we derive the Euclidean path integral representation for the statistical sum of the microcanonical ensemble in cosmology. Cosmology is supposed to be closed with spatial sections of the $S^3$ topology, and the EQG path integration runs over periodic configurations with the topology of $S^3\times S^1$. In the limiting case these configuarations also include the topology of $\bigcup_{i=1}^k S_i^4$.  The latter can be viewed as the result of contracting the $S^3$ section of $S^3\times S^1$ to a zero size -- a point -- at $k$ different locations on $S^1$. This contraction or pinching procedure explains the class of topological transitions we are going to consider here. In more general terms they can be described as taking the connected patch of a manifold (say, between the $S^3$-surfaces $\Sigma$ and $\Sigma'$ as depicted on Fig.\ref{Fig.3}) and pinching it at some point in such a way that it gets disconnected into two parts, this pinching point becoming {\em a regular internal point} of each of these parts.\footnote{This pinching point can also be viewed as shared by these two parts, but then the whole manifold is no longer Hausdorff separable.} A similar transition leading to multiple spheres configurations is depicted on Fig.\ref{Fig.3a} for $k=4$ pinching points. Important remark is that such a Euclidean manifold inherits from the Lorentzian setup a distinguished arrow of time (or, better to say, directionless axis of time -- it acquires direction only after continuation to the Lorentzian spacetime which nucleates from the Euclidean space at minimal surfaces). This arrow of time does not only implement periodicity on the circle $S^1$ of $S^3\times S^1$, but also remains built in every $S_i^4$ of $\bigcup_{i=1}^k S_i^4$ because it runs through the sequence of the pinching points of the above type.

Though it is hard to implement explicitly the regularity of such pinching points in the EQG path integration, this property can be observed in the saddle points of the path integral. This is done in Sects.3 and 4 where according to \cite{slih} the semiclassical calculation of the statistical sum is applied to the CFT driven cosmology. It is shown that the no-boundary instantons of the Hartle-Hawking type originate in this model by such a pinching procedure, but their contribution is dynamically suppressed to zero due to the effect of the conformal anomaly of quantum fields. It is shown, in particular, that this suppression is an artifact of special junction conditions for quantum fields, which in their turn are enforced by the arrow of time inherited from the Lorentzian theory. In Sect.5 we show that the suppression of topology transitions goes beyond CFT models and holds in the one-loop approximation for generic local boson and fermion fields as a general consequence of these junction conditions. In Sect.6 we discuss the problem of equivalence of different dynamical mechanisms of topology change suppression, its extension beyond one-loop order and a number of related issues.

\section{Euclidean quantum gravity from the physical theory in Lorentzian spacetime}
The physical setting in Lorentzian signature spacetime starts with the definition of the microcanonical ensemble in canonically quantized gravity theory. In cosmology the corresponding density matrix $\hat\rho=\rho(\varphi,\varphi')$ was suggested in \cite{why} as a formal projector
    \begin{eqnarray}
    \hat\rho\sim
    \prod_\mu \delta(\hat H_\mu)             \label{projector}
    \end{eqnarray}
on the subspace of physical states satisfying the system of the Wheeler-DeWitt equations
    \begin{eqnarray}
    \hat H_\mu(\varphi,\partial/i\partial\varphi)\, \rho(\varphi,\varphi')=0,                      \label{WDW}
    \end{eqnarray}
where $\hat H_\mu$ denotes the operator realization of the full set of the gravitational Hamiltonian and momentum constraints
$H_\mu(q,p)$. The formal product in (\ref{projector}) runs over the condensed index $\mu$ signifying a collection of discrete labels along with continuous spatial coordinates,
$\mu=(\perp, a, {\bf x}),\,\,a=1,2,3$. The phase space variables $(q,p)$ include the collection of spatial metric coefficients and matter fields $q=(g_{ab}\big({\mathbf x}),\phi({\mathbf x})\big)$ and their conjugated momenta $p$. The canonical coordinates $q$ will be also denoted by $\varphi$ when used as arguments of the density matrix kernel $\langle\varphi|\,\hat\rho\,|\varphi'\rangle
=\rho(\varphi,\varphi')$.

The justification for (\ref{projector}) as the density matrix of a {\em microcanonical} ensemble in spatially closed cosmology was put forward in \cite{why} based on the analogy with an unconstrained system having a conserved Hamiltonian $\hat H$. The microcanonical state with a fixed energy $E$ for such a system is given by the density matrix $\hat\rho\sim \delta(\hat H-E)$. A major distinction of (\ref{projector}) from this case is that spatially closed cosmology does not have freely specifiable constants of motion like the energy or other global charges. Rather it has as constants of motion the Hamiltonian and momentum constraints $H_\mu$, all having
a particular value --- zero. Therefore, the expression
(\ref{projector}) can be considered as the analogue of equipartition -- a natural candidate for the microcanonical quantum state of the {\em closed} Universe.

The definition (\ref{projector}) has, of course, a very formal nature because it is very incomplete in view of non-commutativity of the constraints $\hat H_\mu$, infinite dimensional (and continuous) nature of the space of indices $\mu$ and the phase space, etc. However, at the semiclassical level (within the perturbation loop expansion) the kernel of this projector can be written down as a Faddeev-Popov gauge-fixed path integral of the canonically quantized gravity theory \cite{PhysRep,why}
    \begin{eqnarray}
    \rho(\varphi_+,\varphi_-)=
    e^{\varGamma}\!\!\!\!\!\!\!\!\!\!\!
    \int\limits_{\,\,\,\,\,\,
    q(t_\pm)=\,\varphi_\pm}
    \!\!\!\!\!\!\!\!\!
    D[\,q,p,N\,]\;
    \exp\left[\,i\!
    \int_{\,\,t_-}^{t_+} dt\,
    (p\,\dot q-N^\mu H_\mu)\,\right].   \label{rhocanonical}
    \end{eqnarray}
Here $N^\mu$ are the Lagrange multipliers dual to the constraints -- lapse and shift functions $N^\mu=(N({\mathbf x}),N^a({\mathbf x}))$, and the functional integration runs over the histories interpolating between the configurations $\varphi_\pm$ which are the arguments of the density matrix kernel. The range of integration over $N^\mu$ is
of course real because this integration over the Lagrange
multipliers is designed in order to generate delta functions of constraints. The Hamiltonian action in the exponential is the
integral over the coordinate time $t$ which is just the ordering parameter ranging between arbitrary initial and final values $t_\pm$, the result being entirely independent of their choice.\footnote{The projector on the space of non-commuting constraints can be realized by integration over the relevant group. Integration over the canonically realized diffeomorphisms implicit in the gauge fixed integral over $N^\mu$  is just the analogue of this group integration. On the other hand, the chronological ordering generated by the path integration in (\ref{rhocanonical}) takes care of the operator ordering in (\ref{projector}). Since in closed cosmology there is no non-vanishing Hamiltonian, the history parameter $t$ in (\ref{rhocanonical}) exclusively serves this operator-ordering role. This is a peculiarity of the theories with a parameterized time \cite{PhysRep}.} The integration measure $D[\,q,p,N\,]$, of course, includes the Faddeev-Popov gauge-fixing procedure which renders the whole integral gauge independent.\footnote{Original definition (\ref{projector}) might seem contradictory because equipartition over entire physical space, not restricted by a fixed value of some observable like energy, is likely to result in divergent statistical sum and expectation values. However, the path integral representation (\ref{rhocanonical}) treated within $\hbar$-expansion implies projection onto perturbative solutions of the Wheeler-DeWitt equations (\ref{WDW}), which have a semiclassical limit. As we will see below, in the model of interest this indeed leads to saddle-point approximation with finite characteristics and, moreover, unexpectedly provides their limited range.}

After integration over canonical momenta the path integral above takes the Lagrangian form of the integral over the configuration space coordinates $q$ and the lapse and shift functions $N^\mu$. Taken together they comprise the full set of the spacetime metric with the Lorentzian signature $g_{\mu\nu}^L$ and matter fields $\phi$ ,
    \begin{eqnarray}
    g_{\mu\nu}^Ldx^\mu dx^\nu=-N^2_L dt^2+g_{ab}(dx^a+N^a dt)(dx^b+N^b
    dt),
    \end{eqnarray}
in terms of which the Lagrangian form of the classical action reads as $S[\,g_{\mu\nu}^L,\phi\,]$. One more notational
step consists in the observation that this Lorentzian metric can be viewed as the Euclidean metric $g_{\mu\nu}$ with the imaginary value of the Euclidean lapse function $N$,
    \begin{eqnarray}
    &&g_{\mu\nu} dx^\mu dx^\nu=N^2 d\tau^2
    +g_{ab}(dx^a+N^a d\tau)\,(dx^b+N^b
    d\tau),\\                          \label{decomposition}
    &&N=iN_L,                               \label{NLvsNE}
    \end{eqnarray}
so that the Euclidean theory action is related to the original
Lorentzian action $S_L[\,g_{\mu\nu}^L,\phi\,]$ by a typical equation
    \begin{eqnarray}
    i S_L[\,g_{\mu\nu}^L,\phi\,]
    =-S[\,g_{\mu\nu},\phi\,].               \label{ELactions}
    \end{eqnarray}
Here the imaginary factor arises from the square root of the metric determinant in the Lagrangian, which in the ADM form reads as $g^{1/2}=N(\det g_{ab})^{1/2}$. Note that the analytic continuation from the Lorentzian to the Euclidean picture takes place in the complex plane of the lapse function rather than in the complex plane of time (time variable is the same in both pictures  $\tau=t$), though of course it is equivalent to the usual Wick rotation.

With these notations the density matrix (\ref{rhocanonical}) takes the form of the Euclidean quantum gravity path integral
    \begin{eqnarray}
    \rho(\,\varphi_+,\varphi_-\,)=
    \mbox{\large$e$}^{\,\varGamma}\!\!\!\!\!\!\!\!\!
    \int\limits_{\,\,\,\,\,\,
    q(t_\pm)=\,\varphi_\pm}
    \!\!\!\!\!\!\!
    D[\,g_{\mu\nu},\phi\,]\,
    e^{-S[\,g_{\mu\nu},\phi\,]}.             \label{rho}
    \end{eqnarray}
However, in view of (\ref{NLvsNE}) the range of integration over the Euclidean lapse $N$ belongs to the imaginary axis
    \begin{eqnarray}
    -i\infty<N<i\infty.  \label{imaginaryaxis}
    \end{eqnarray}
\begin{figure}[h]
\centerline{\includegraphics[width=10cm]{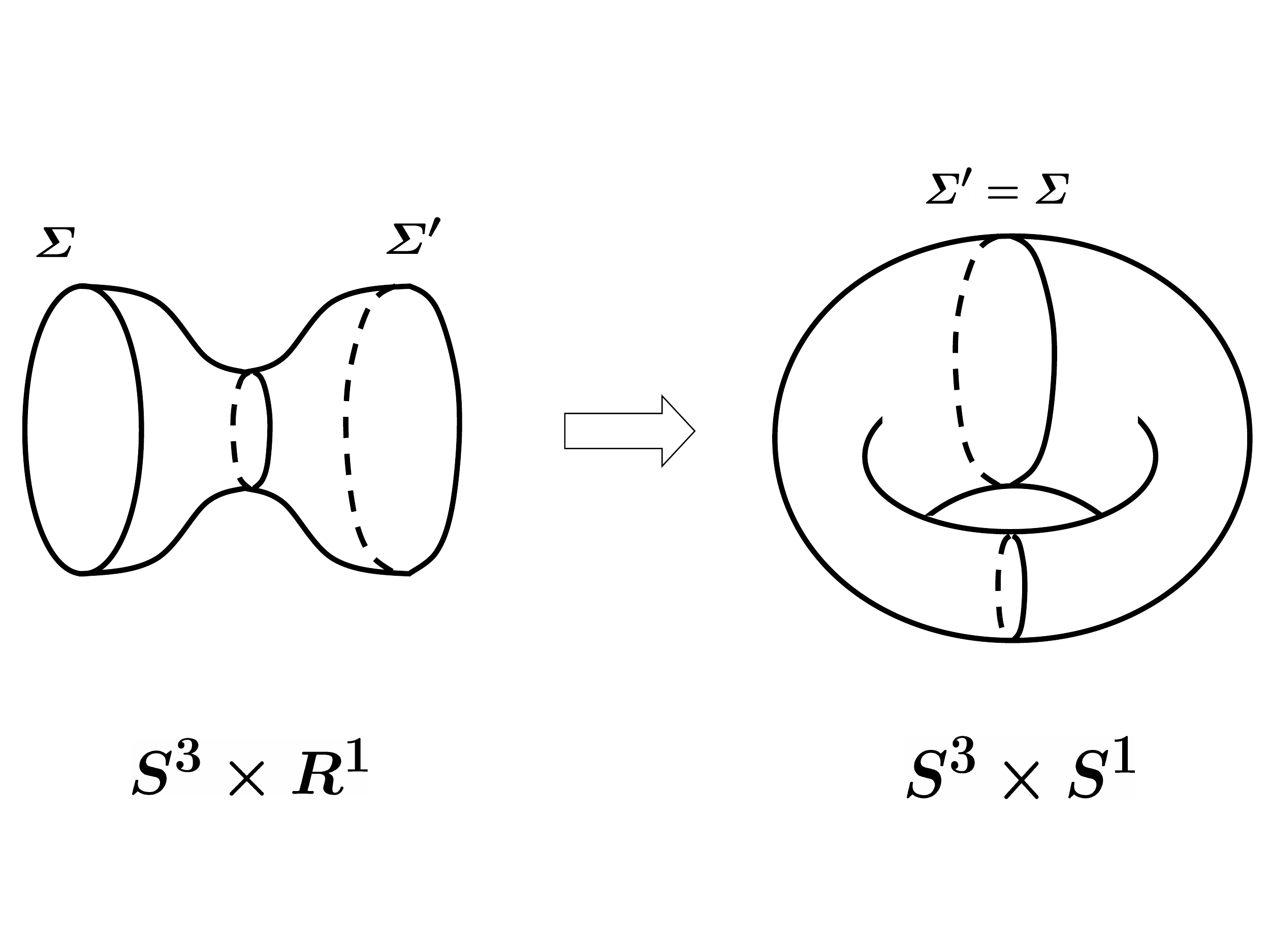}} \caption{\small
Transition from the density matrix to the statistical sum.
 \label{Fig.2}}
\end{figure}

The topology of spacetime configurations which are integrated over in (\ref{rho}) is $S^3\times R^1$ as depicted on the left part of Fig.\ref{Fig.2}. This topology of the spacetime bulk interpolating between the hypersurfaces $\Sigma$ and $\Sigma'$ reflects the mixed nature of the density matrix and establishes entanglement correlations between $\varphi$ and $\varphi'$ \cite{slih}. These configurations however include as a limiting case the bulk obtained by pinching the spacetime bridge between $\Sigma$ and $\Sigma'$ (see Fig.\ref{Fig.3}) -- the bulk spacetime decomposes into two parts having in common one single point which is a regular internal point to each of these two pieces.  The contribution of this configuration was associated in \cite{slih,why} with the direct product of pure states of the Hartle-Hawking type. Below we will see that this interpretation is not quite precise. While this interpretation seems correct in the tree-level approximation, beyond it quantum fluctuations destroy pure state coherence and, in fact, suppress to zero the contribution of these nontrivial topologies. Additional remark is that pinching the spacetime between $\Sigma$ and $\Sigma'$ can take place several times $k=2,3,...$, and the picture of Fig.\ref{Fig.3} should be modified accordingly.

\begin{figure}[h]
\centerline{\includegraphics[width=12cm]{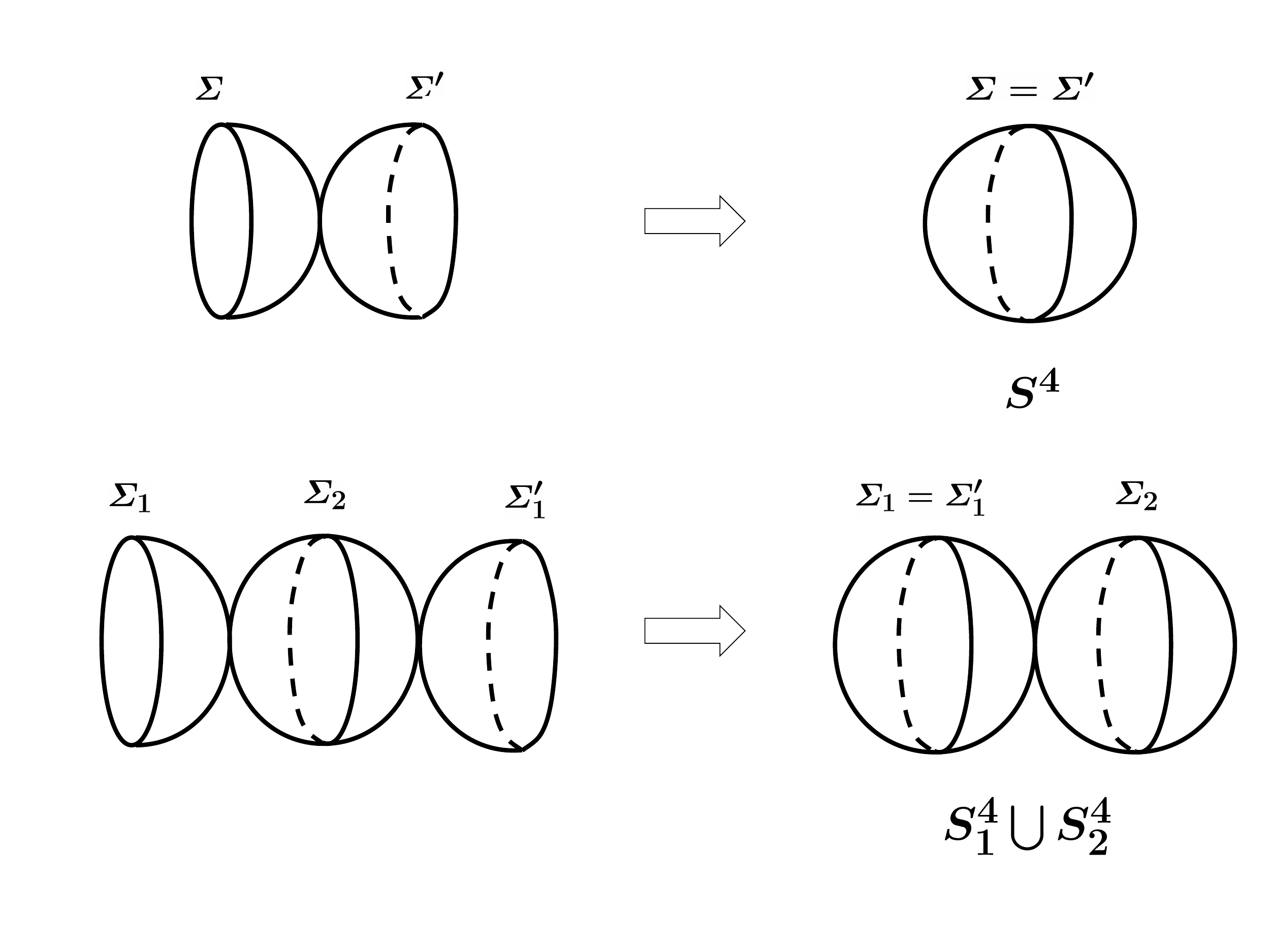}} \caption{\small
Transition from the density matrix to the statistical sum resulting in the no-boundary instantons $\bigcup_{i=1}^k S_i^4$ for $k=1$ and $k=2$.
\label{Fig.4}}
\end{figure}

The normalization factor $\exp\varGamma$ in (\ref{rho}) follows from the density matrix normalization ${\rm tr}\hat\rho=1$ and determines the main object of interest -- the statistical sum of the model. The trace operation implies integration over the diagonal elements of the density matrix, so that the statistical sum takes the form of the path integral
    \begin{eqnarray}
    &&e^{-\varGamma}=
    \!\!\int\limits_{\,\,\rm periodic}
    \!\!\!\! D[\,g_{\mu\nu},\phi\,]\;
    e^{-S [\,g_{\mu\nu},\phi\,]}   \label{EuclideanPI}
    \end{eqnarray}
over periodic configurations whose spacetime topology $S^3\times S^1$ follows from the identification of the boundary surfaces $\Sigma$ and $\Sigma'$. This leads to to the topology $S^1\times S^3$ depicted on the right part of Fig.\ref{Fig.2}, whereas the ``pure'' state contribution yields the topology which in the simplest case of one pinch is $S^4$ as depicted on the upper part of Fig.\ref{Fig.4}. For multiple pinches the topology becomes $\bigcup_{i=1}^k S_i^4$ as shown on the lower part of this figure for $k=2$.

Formally the Euclidean path integral (\ref{EuclideanPI}) is independent of a particular foliation of the spacetime by the $(3+1)$-decomposition (\ref{decomposition}), which is guaranteed by the Faddeev-Popov gauge fixing procedure implicit in the integration measure \cite{PhysRep}. This procedure provides gauge independence of (\ref{EuclideanPI}) with respect to {\em local} changes of the coordinate gauge. However, the imaginary lapse integration range (\ref{imaginaryaxis}) and periodicity of $\tau\in S^1$ have a global or topological nature. They select a distinguished arrow of time which is inherited in the Euclidean formalism from the physical setting in the Lorentzian spacetime. As we will see below, this arrow of time leads to certain boundary or junction conditions for quantum fields, which suppress the contribution of nontrivial topologies to the statistical sum.

\section{CFT driven cosmology and its thermal states}
The actual calculation of the statistical sum can be based on decomposing the full set of fields, $[\,g_{\mu\nu}(x),\phi(x)\,]\to [\,a(\tau),N(\tau);\,\varPhi(x)\,]$, into the minisuperspace sector of the spatially closed Friedmann-Robertson-Walker (FRW) metric,
    \begin{eqnarray}
    g_{\mu\nu}^{FRW} dx^\mu dx^\nu=N^2(\tau)\,d\tau^2
    +a^2(\tau)\,d^2\Omega^{(3)}, \label{FRW}
      \end{eqnarray}
and all spatially inhomogeneous ``matter'' fields $\varPhi(x)=\varPhi(\tau,{\bf x})$, $\varPhi(x)=[\,\phi(x),\psi(x),A_\mu(x), h_{\mu\nu}(x),...\,]$. Then the path integral can be cast into the form of an integral over a minisuperspace lapse function $N(\tau)$ and scale factor $a(\tau)$ of this metric,
    \begin{eqnarray}
    &&e^{-\varGamma}=\int
     D[\,a,N\,]\;
    e^{-S_{\rm eff}[\,a,\,N\,]},              \label{1}\\
    &&e^{-S_{\rm eff}[\,a,\,N]}
    =\int D[\,\varPhi(x)\,]\,
    e^{-S[\,a,\,N;\,\varPhi(x)\,]},             \label{2}
    \end{eqnarray}
where $S_{\rm eff}[\,a,\,N\,]$ is the effective action of all these ``matter'' fields $\varPhi(x)$ (which include also the metric perturbations $h_{\mu\nu}(x)$) on the minisuperspace background of the FRW metric. The action $S[a,N;\varPhi(x)]\equiv S[\,g_{\mu\nu},\phi\,]$ is the original Euclidean action rewritten in terms of the above minisuperspace decomposition.

This construction has a predictive power in the gravitational model with a matter sector dominated by a large number of free (linear) fields conformally coupled to gravity -- conformal field theory (CFT)
    \begin{eqnarray}
    &&S[\,g_{\mu\nu},\phi\,]=-\frac1{16\pi G}
    \int d^4x\,g^{1/2}\,(R-2\varLambda)
    +S_{CFT}[\,g_{\mu\nu},\phi\,].
    \end{eqnarray}
The effective action in such a system is dominated by the quantum action of these conformal fields which simply outnumber the non-conformal fields (including the graviton),
    \begin{eqnarray}
    &&S_{\rm eff}[\,a,\,N\,]\simeq\left(\,-\frac1{16\pi G}
    \int d^4x\,g^{1/2}\,(R-2\varLambda)
    +S_{CFT}^{\rm eff}[\,g_{\mu\nu}]\,\right)
    _{\;g_{\mu\nu}=g_{\mu\nu}^{FRW}},\\
    &&\mbox{\large$e$}^{\,-S_{CFT}^{\rm eff}[\,g_{\mu\nu}]}=
    \int D[\,\phi\,]\,
    e^{-S_{CFT}[\,g_{\mu\nu},\phi\,]}
    =\left({\rm Det}\,
    \frac{\delta^2 S_{CFT}}{\delta\phi(x)\,
    \delta\phi(y)}\right)^{-1/2}.          \label{det}
    \end{eqnarray}
This quantum effective action, in its turn, is exactly calculable by the conformal transformation converting (\ref{FRW}) into the static Einstein metric with $a={\rm const}$. It becomes the sum of the contribution of this conformal transformation \cite{FHHS,anomalyaction}, determined by the well-known conformal anomaly of a quantum CFT in the external gravitational field \cite{Duffanomaly} and the contribution of a static Einstein Universe -- the combination of the Casimir energy \cite{Casimir} and free energy of a typical boson or fermion statistical sum.  The temperature of the latter is given by the inverse of the Euclidean time period of the $S^1\times S^3$ instanton, measured in units of the conformal time.

Namely, this effective action reads in units of the Planck mass $m_P=(3\pi/4G)^{1/2}$ \cite{slih}
    \begin{eqnarray}
    &&S_{\rm eff}[\,a,N\,]=m_P^2\int_{S^1} d\tau\,N \left\{-aa'^2
    -a+ \frac\varLambda3 a^3+\,B\left(\frac{a'^2}{a}
    -\frac{a'^4}{6 a}\right)+\frac{B}{2a}\,\right\}
    +F(\eta),                              \label{effaction}\\
    &&F(\eta)=\pm\sum_{\omega}\ln\big(1\mp
    e^{-\omega\eta}\big),                 \label{freeenergy}\\
    &&\eta=\int_{S^1} \frac{d\tau N}a,     \label{period}
    \end{eqnarray}
where $a'\equiv da/Nd\tau$. The first three terms in curly brackets of (\ref{effaction}) represent the Einstein action with a primordial (but renormalized by quantum corrections) cosmological constant $\varLambda\equiv 3H^2$ ($H$ is the corresponding Hubble constant). The terms proportional to the constant $B$ correspond to the contribution of the conformal anomaly and the contribution of the vacuum (Casimir) energy $(B/2a)$ of conformal fields on a static Einstein spacetime of the size $a$. Finally, $F(\eta)$ is the free energy of these fields -- a typical boson or fermion sum over CFT field oscillators with energies $\omega$ on a unit 3-sphere, $\eta$ playing the role of the inverse temperature --- an overall circumference of the $S^1\times S^3$ instanton in terms of the conformal time (\ref{period}).

The constant $B$,
    \begin{eqnarray}
    B=\frac{3\beta}{4 m_P^2},         \label{B}
    \end{eqnarray}
is determined by the coefficient $\beta$ of the topological Gauss-Bonnet invariant $E=R_{\mu\nu\alpha\gamma}^2 -4R_{\mu\nu}^2 + R^2$ in the overall conformal anomaly
    \begin{eqnarray}
    g_{\mu\nu}\frac{\delta
    S^{CFT}_{\rm eff}}{\delta g_{\mu\nu}} =
    \frac{1}{4(4\pi)^2}g^{1/2}
    \left(\alpha \Box R +
    \beta E +
    \gamma C_{\mu\nu\alpha\beta}^2\right), \label{anomaly}
    \end{eqnarray}
which is always positive for any CFT particle content \cite{Duffanomaly}. The UV ambiguous coefficient $\alpha$ was renormalized  to zero by a local counterterm $\sim \alpha R^2$ to guarantee the absence of higher derivative terms in the action (\ref{effaction}). This automatically gives the renormalized Casimir energy the value $m_P^2B/2a=3\beta/8a$ which universally expresses in terms of the same coefficient in the conformal anomaly \cite{universality}.\footnote{This
universality property follows from the fact that in a static Einstein Universe of the size $a$ the Casimir energy of conformal fields is determined by the conformal anomaly coefficients and equals $(3\beta-\alpha/2)/8a$ \cite{universality}.} The coefficient $\gamma$ of the Weyl tensor squared term $C^2_{\mu\nu\alpha\beta}$ does not enter the expression (\ref{effaction}) because $C_{\mu\nu\alpha\beta}$ identically vanishes for any FRW metric.

Semiclassically the statistical sum (\ref{1}) is dominated by the solutions of the effective equation for the action (\ref{effaction}), $\delta S_{\rm eff}/\delta N(\tau)=0$. This is the modification of the Euclidean Friedmann equation,
    \begin{eqnarray}
    &&-\frac{a'^2}{a^2}+\frac{1}{a^2}
    -B \left(\,\frac{a'^4}{2a^4}
    -\frac{a'^2}{a^4}\right) =
    \frac\varLambda3+\frac{C}{ a^4},     \label{efeq}\\
    &&C = \frac{B}2+\frac1{m_P^2}\frac{dF(\eta)}{d\eta}=
    \frac{B}2+
    \frac1{m_P^2}
    \sum_\omega\frac{\omega}
    {e^{\omega\eta}\mp 1},               \label{bootstrap}
    \end{eqnarray}
by the anomalous $B$-term and the radiation term $C/a^4$ with the constant $C$ characterizing the sum of the Casimir energy and the energy of the gas of thermally excited particles with the inverse temperature $\eta$ given by (\ref{period}).

As shown in \cite{slih,why,DGP/CFT} the solutions of this integro-differential equation\footnote{Note that the constant $C$ is a nonlocal functional of the history $a(\tau)$ -- Eq.(\ref{bootstrap}) plays the role of the bootstrap equation for the amount of radiation determined by the background on top of which this radiation evolves and produces back reaction.} give rise to two types of instantons. The first type represents the set of periodic $S^3\times S^1$ instantons with the oscillating scale factor -- {\em garlands} (cf. Fig.\ref{Fig.3a}) that can be regarded as the thermal version of the Hartle-Hawking instantons. In these solutions the scale factor oscillates $k$ times ($k=1,2,3,...$) between its maximum and minimum values
$a_\pm=a(\tau_\pm)$, $a_-\leq a(\tau)\leq a_+$,
    \begin{eqnarray}
    a^2_\pm=\frac{1\pm\sqrt{1-4H^2C}}{2H^2},     \label{apm}
    \end{eqnarray}
so that the full period of the conformal time (\ref{period}) is given by the $2k$-multiple of the integral between the two neighboring turning points of the scale factor history $a(\tau)$, $\dot a(\tau_\pm)=0$,
    \begin{eqnarray}
    &&\eta=2k\int_{\tau_-}^{\tau_+}
    \frac{d\tau\,N}{a}=2k\int_{a_-}^{a_+}
    \frac{da}{a'a}.                       \label{period1}
    \end{eqnarray}
This value of $\eta$ is finite and determines a finite effective temperature $T=1/\eta$ as a function of $G$ and $\varLambda$. According to Sect.2 this is the artifact of a microcanonical ensemble in cosmology \cite{why} with only two freely specifiable dimensional parameters --- the renormalized gravitational and renormalized cosmological constants.

These $S^3\times S^1$ garland-type instantons exist only in the limited range of the cosmological constant $\varLambda=3H^2$ \cite{slih},
    \begin{eqnarray}
    0<\varLambda_{\rm min}<\varLambda<
    \varLambda_{\rm max}=\frac3{2B},                \label{landscape}
    \end{eqnarray}
where the spectrum of admissible values of $\varLambda$ has a band structure. The countable ($k=1,2,3,...$) sequence of bands $\Delta_k$ of ever narrowing widths, each of them corresponding to k-fold instantons of the above type, with $k\to\infty$ accumulates at the upper bound of this range. Periodicity of all these instantons originates from the tracing operation signifying the transition from the density matrix to the statistical sum, which is depicted on Fig.\ref{Fig.2} for the one-folded case $k=1$.

\section{The no-boundary states and the conformal mechanism of their dynamical suppression}
The second type of solutions in the CFT driven cosmology is the set of vacuum Hartle-Hawking instantons with the $S^4$-topology and the topology of multiple 4-spheres $\bigcup_{i=1}^k S_i^4$. The existence of these solutions follows from the fact that Eq.(\ref{efeq}) can be rewritten in the following form with $a$-dependent effective Hubble factors $H_{\pm}(a)$,
    \begin{eqnarray}
    &&a'^2=1-a^2\,H^2_{\pm}(a),         \label{efeq1}\\
    &&H^2_{\pm}(a)\equiv
    \frac1B\,\left(1\pm\sqrt{1-2BH^2-\frac{B(2C-B)}{a^4}}\right).
    \end{eqnarray}
In the case of $C=B/2$ these factors become constant and yield as solutions two exact Euclidean de Sitter spacetimes with the effective Hubble constants
    \begin{eqnarray}
    H^2_{\pm}=\frac{1\pm\sqrt{1-2BH^2}}B.   \label{Hpm}
    \end{eqnarray}
The scale factor $a(\tau)$ ranges from zero to the maximum value $a_{\rm max}$ given by their respective turning points $a_{\rm max}=a_\mp=1/H_\pm$ coinciding with (\ref{apm}) at $C=B/2$. This value of the constant $C$ is consistent with the bootstrap equation (\ref{bootstrap}) because the total period of the conformal time instead of Eq.(\ref{period}) is given now by the integral
    \begin{eqnarray}
    &&\eta=2\int\limits_{0}^{a_{\rm max}}
    \frac{da}{a'a}=\infty                 \label{period2}
    \end{eqnarray}
divergent at the lower limit. The corresponding temperature turns out to be zero and the total amount of radiation constant (\ref{bootstrap}) reduces to the contribution of the Casimir energy, which altogether justifies the interpretation of these instantons as the vacuum ones. In contrast to thermal $S^1\times S^3$ instantons belonging to the band spectrum of admissible $\varLambda$, these vacuum instantons exist for all possible $\varLambda\leq 3/2B$ (or $1-2BH^2\geq 0$ when the expressions (\ref{Hpm}) for $H_\pm$ make sense).

Topologically they correspond to a Euclidean hemisphere with a regular internal point $a=0$ without a conical singularity, because in view of (\ref{efeq1}) $a'=1$ at this point. Therefore this is a situation of the no-boundary instantons of the Hartle-Hawking type, though graphically the density matrix origin of this construction suggests two hemispheres glued together at their poles. Their origin can be qualitatively depicted as pinching the segment of the $S^3\times S^1$ instanton at some point as it is shown on Fig.\ref{Fig.3} for one pinch and on Fig.\ref{Fig.3a} for four pinching points.\footnote{These figures should not be interpreted  literally as the transition from garlands to multiple spheres, because these pinches occur not as a zero limit $a_-\to 0$ of the minimum value of $a(\tau)$ in the garland solution. Rather, in the case of $C=B/2$ the turning points $a_\pm$  respectively for the two de Sitter solutions with the Hubble constants $H_\mp$ both serve as the maximum values of their $a(\tau)$. The domain of the Euclidean evolution with $a'^2\geq 0$ lies in this case not between $a_-$ and $a_+$, but belongs to the range $0\leq a\leq a_{\rm max}$.} The transition from the density matrix to the statistical sum for these no-boundary instantons looks different from Fig.\ref{Fig.2} and obviously gives multiple spheres instantons $\bigcup_{i=1}^k S_i^4$ depicted on Fig.\ref{Fig.4} for $k=1$ and $k=2$.

Both thermal and vacuum instantons are semiclassically weighted by the exponentiated onshell value of the action (\ref{effaction}), $\exp(-\varGamma_0)$, which reads \cite{slih}
    \begin{equation}
    \varGamma_0= F(\eta)\!-\!\eta \frac{dF(\eta)}{d\eta}
    +2m_P^2\int_{S^1} d\tau\,N\,
    \frac{a'^2}{a}\Big(B-a^2
    -\frac{Ba'^2}{3}\Big),        \label{onshell}
    \end{equation}
where the integration runs over the full period of the instanton time. It is finite for thermal instantons, but diverges to $+\infty$ for the vacuum instantons of the Hartle-Hawking type with $F\sim dF/d\eta=0$,
    \begin{equation}
    \varGamma_0\,\Big|_{\; HH}=
    4km_P^2\int\limits_0^{a_{\rm max}} \frac{da}a\,
    a'\,\Big(B-a^2
    -\frac{Ba'^2}{3}\Big)\sim
    \frac{8k m_P^2 B}3\int\limits_0^{a_{\rm max}}
    \frac{da}a\to+\infty,          \label{onshell1}
    \end{equation}
because of the logarithmic divergence at $a=0$. The latter is contributed by $k$ points with a vanishing cosmological scale factor (defined in the foliation transversal to the arrow of time which was inherited in the Euclidean formalism from its Lorentzian counterpart). Since $B>0$ for conformal particles of all low order spins (see Eq.(\ref{100}) below), this divergence completely rules out vacuum instantons and leaves in the statistical sum only thermal contributions.

This is a combined effect of the conformal anomaly and Casimir energy -- their contribution to (\ref{effaction}) diverges at the points with $a(\tau)=0$. It drastically changes the predictions of the tree-level approximation characterized by a negative onshell action $\sim -1/H^2$ and, thus, eliminates a very anti-intuitive situation of infinitely enhanced creation of infinitely large universes with $H\to 0$. However, this conclusion sounds paradoxical because it would mean a divergent value of the functional determinant (\ref{det})
    \begin{eqnarray}
    &&\mbox{\large$e$}^{\,-\varGamma_0}\sim
    \left({\rm Det}\,
    \frac{\delta^2 S_{CFT}}{\delta\phi(x)\,
    \delta\phi(y)}\right)^{-1/2}          \label{det1}
    \end{eqnarray}
calculated on regular spherical instantons $\bigcup_{i=1}^k S_i^4$ of Fig.\ref{Fig.3}. This determinant should of course be finite after the UV renormalization (which has already been implicitly done when deriving the expression (\ref{effaction}) for the effective action and is responsible, in particular, for its conformal anomaly contribution).\footnote{I am grateful to V.Rubakov for pointing out to this paradox, which eventually has led to a completion of this work.}

The resolution of this paradox consists in the observation that the pictures of Fig.\ref{Fig.3}, Fig.\ref{Fig.3a} and Fig.\ref{Fig.4} do not reflect properly boundary conditions on the integration fields in the path integral (or the space of functions on which the functional determinant in (\ref{det1}) is defined). These boundary conditions are missing in the naive EQG formulation which only resorts to the regularity of fields on a smooth compact Euclidean spacetime without boundary. Alternatively, in the Lorentzian-Euclidean setup of Sect.2 these special boundary conditions are inherited in the Euclidean formalism from the Lorentzian side with its inherent arrow of time.
\begin{figure}[h]
\centerline{\includegraphics[width=10cm]{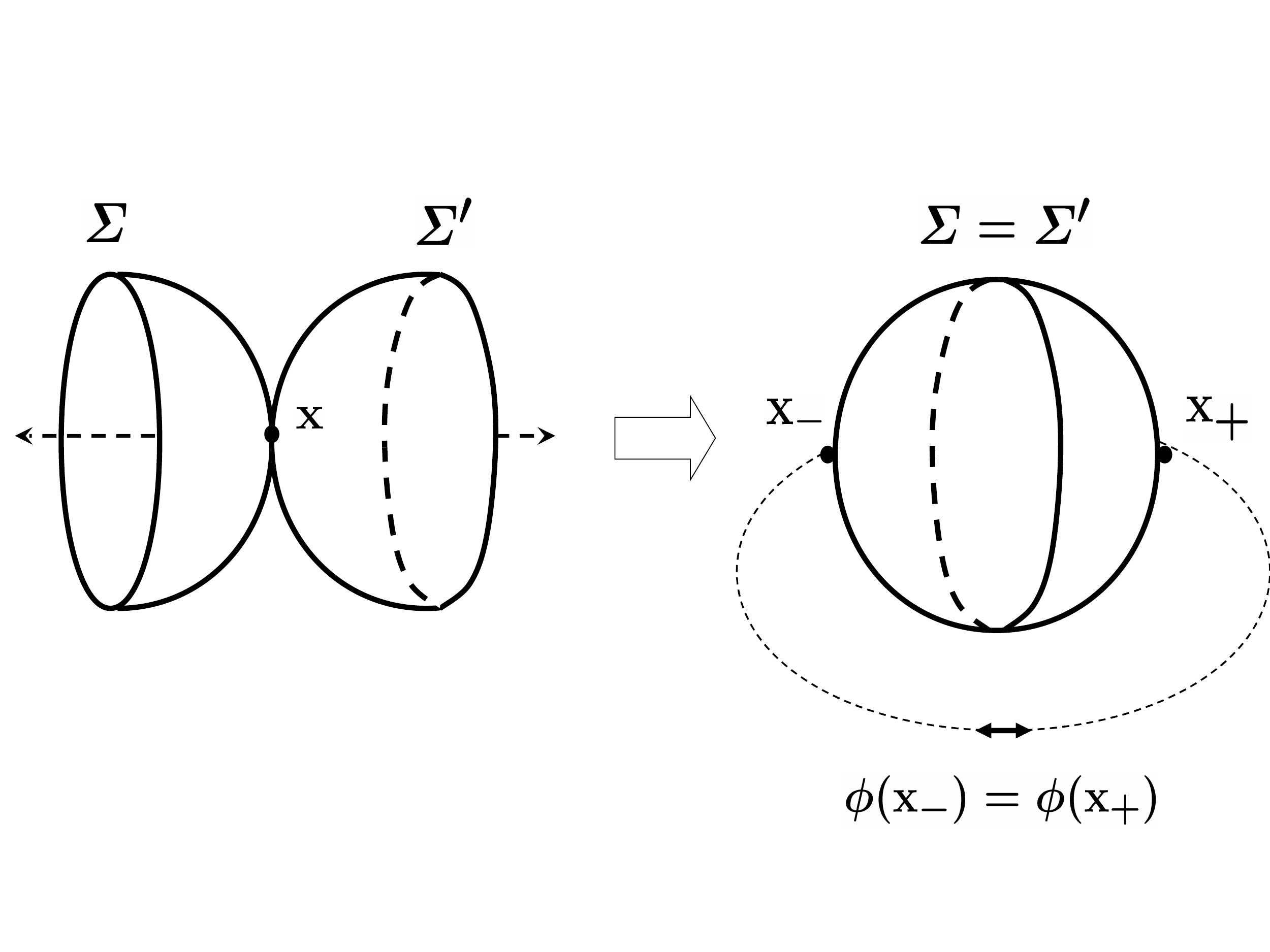}} \caption{\small
Origin of junction conditions: in the transition to the statistical sum the pinching point $x$ goes over into two different points of $S^4$, $x\to x_\pm$, with equal field values and the arrow of time running through them.
 \label{Fig.5}}
\end{figure}

The origin of these boundary conditions can be explained by Fig.\ref{Fig.5}. In the path integral for the density matrix the 4-dimensional integration field continuously interpolates between 3-dimensional configurations on $\varSigma$ and $\varSigma'$. Therefore, the values of this field at the junction of the two spacetime hemispheres should match when approaching this junction point either from the left or from the right hemisphere. In the transition to the statistical sum this junction point splits into two different points $x_\pm$ on the resulting $S^4$ instanton, so that quantum fields on this instanton satisfy the boundary (or, better to say, junction) condition
    \begin{eqnarray}
     \phi(x_-)=\phi(x_+).    \label{junction}
    \end{eqnarray}
The distinguished role of these two points on a smooth sphere $S^4$ follows from the fact that they lie on a (double-sided) arrow of time inherited on the density matrix instanton from the definition of this matrix in the physical Lorentzian spacetime (cf. an alternative suggestion of the arrow of time imposed on the Lorentzian theory from spherically asymmetric nucleation of branes in Euclidean theory \cite{Gorsky}).

The calculation of the functional determinant (\ref{det1}) by the method of the conformal transformation from a non-static $S^1\times S^3$ spacetime to a static Einstein Universe implicitly incorporates these junction conditions when $S^1\times S^3$ degenerates to $S^4$. Indeed, the conformal factor $\sim a(\tau)$ relating the metrics of the spaces on the upper and lower parts of Fig.\ref{Fig.3} gets singular at the junction point of two hemispheres ($a=0$), and thus leads to the result (\ref{onshell1}) diverging to $+\infty$. At the same time, the functional integration runs over continuous fields, and in the limit when the manifold gets pinched at $a=0$ the fields on the poles of these hemispheres remain identified even though these poles can be treated as disconnected. Thus the continuity of fields in the path integral enforces the junction condition (\ref{junction}), and the functional integration does not run independently on the smooth caps of these hemispheres. Equivalently, in the statistical sum, reconnecting these hemispheres into one $S^4$, the functional determinant (\ref{det1}) is defined on the space of functions subject to (\ref{junction}) rather than on the space of all regular functions on $S^4$. This explains the difference between these two determinants which are related by the divergent factor contributed by (\ref{onshell1}).

\section{General mechanism of dynamical suppression}
In fact, dynamical suppression of topological transitions of the form depicted on Fig.\ref{Fig.3} has a more general nature and does not rely on conformal properties of the quantum field. Generically the topology change of Fig.\ref{Fig.3} implies that a connected patch of the full manifold goes over into two smooth disconnected patches having two regular internal points $x_\pm$ at which the values of the field are identified as in (\ref{junction}). Quantum fluctuations subject to such junction conditions suppress to zero the contribution of such spacetime manifolds. For local boson fields this happens due to their UV behavior, whereas for fermions this is a simple corollary of the Pauli principle. This looks as follows.

For any boson field $\phi(x)$ with a local quadratic action of the form
    \begin{eqnarray}
     S[\,\phi\,\,]=\frac12\int
     dx\,\phi(x)F(\nabla)\phi(x),
    \end{eqnarray}
the statistical sum over configurations with identified fields at two different spacetime points $x_\pm$ reads as
    \begin{eqnarray}
    &&Z=\!\!\int\limits_{\phi(x_+)=\phi(x_-)}\!\! D\phi\,\exp\big(-S[\,\phi\,\,]\big).
    \end{eqnarray}
It can be rewritten in terms of the delta function in the integrand, which in its turn can be represented as an additional integral over the numerical (not functional) integral over the Lagrange multiplier $\lambda$. The subsequent Gaussian integration over $\phi(x)$ then gives
    \begin{eqnarray}
    &&Z=\int D\phi\,\exp\big(-S[\,\phi\,\,]\big)\,
    \delta\Big(\phi(x_+)-\phi(x_-)\Big)\nonumber\\
    &&\qquad\qquad=
    \int D\phi\;d\lambda\,\exp\Big(-S[\,\phi\,\,]
    +i\lambda\big(\phi(x_+)-\phi(x_-)\big)\Big)\nonumber\\
    &&\qquad\qquad=Z_{\rm E}\int d\lambda\,\exp\left(-\frac12\int dx\,dy\,J_\lambda(x)G(x,y)
    J_\lambda(y)\right).            \label{1000}
    \end{eqnarray}
Here $Z_{\rm E}$ is the EQG statistical sum on a smooth manifold -- the path integral over regular fields without any junction conditions, which is perfectly finite after the relevant UV renormalization,
    \begin{eqnarray}
    Z_{\rm E}=\Big({\rm Det} \,F(\nabla)\Big)^{-1/2}.
    \end{eqnarray}

The effect of these junction conditions -- identification of integration fields at $x_\pm$ -- is represented in (\ref{1000}) in terms of $G(x,y)$ which is the Green's function of the field $\phi(x)$
    \begin{eqnarray}
    F(\nabla)\,G(x,y)=\delta(x-y)
    \end{eqnarray}
and $J_\lambda(x)$ -- a special source parametrically depending on $\lambda$ of the form
    \begin{eqnarray}
    &&J_\lambda(x)=\lambda\,\big(\delta(x-x_+)-\delta(x-x_-)\big).
    \end{eqnarray}

Integration over this Lagrange multiplier immediately gives infinitely suppressing factor
    \begin{eqnarray}
    Z=Z_{\rm E}\,\Big(G(x_+,x_+)
    +G(x_-,x_-)-2G(x_+,x_-)\Big)^{-1/2}=0  \label{suppression}
    \end{eqnarray}
due to the ultraviolet behavior of the Green's function at coincident points $G(x_+,x_+)=\infty$ and $G(x_-,x_-)=\infty$ and finiteness of $G(x_+,x_-)$ with $x_+\neq x_-$ in a Euclidean spacetime. This result is independent of the concrete properties of any set of {\em local} boson fields and universally applies in the one-loop approximation.

For fermion fields the mechanism of suppression is even simpler and turns out to be a consequence of the Pauli principle. In this case the usual Dirac action
    \begin{eqnarray}
     S[\,\psi,\bar\psi\,]=\int
     dx\,g^{1/2}\bar\psi(x)\big(i\gamma^\mu\nabla_\mu-m\big)\psi(x)
    \end{eqnarray}
has the first-order form with $\bar\psi$ playing the role of the canonical momentum conjugated to $\psi$. Therefore in the composition law for the kernels of unitary evolution, say, from $\Sigma_1$ to $\Sigma$ and from $\Sigma$ to $\Sigma_2$ only the values of $\psi$ should be matched at $\Sigma$ in their respective path integrals.\footnote{In full accordance with the number of boundary data dictated by the first order in derivatives of the Dirac operator.} When the surface $\Sigma$ degenerates to a point associated with the two points $x_\pm$ on a smooth manifold, the relevant junction condition involves only $\psi(x_\pm)$ but leaves $\bar\psi(x_+)$ and $\bar\psi(x_-)$ unidentified, and the statistical sum takes the form
    \begin{eqnarray}
    &&Z=\!\!\int\limits_{\psi(x_+)=\psi(x_-)}\!\! D\psi\,D\bar\psi\,\exp\big(-S[\,\psi,\bar\psi\,]\big).
    \end{eqnarray}
Similarly to the bosonic case this introduces the integration over the source $\bar J_\eta$ dual to $\psi$, but keeps the source dual to $\bar\psi$ vanishing, $J_\eta=0$, so that the result is vanishing in view of the grassman nature of $\bar\eta$,
    \begin{eqnarray}
    &&Z=\int d\bar\eta \int D\psi\,D\bar\psi\,
    \exp\Big(-S[\,\psi,\bar\psi\,]
    +i\bar\eta\big(\psi(x_+)-\psi(x_-)\big)\Big)\nonumber\\
    &&\qquad\qquad=Z_{\rm E}\int d\bar\eta\,\exp\left(-\int dx\,dy\,\bar J_\eta(x)G(x,y)J_\eta(y)\,
    \Big|_{\,J_\eta=0}\right)
    =Z_{\rm E}\int d\bar\eta=0.     \label{Pauliprinciple}
    \end{eqnarray}
Here of course $Z_{\rm E}=[{\rm Det}\,(i\gamma^\mu\nabla_\mu-m\big)]^{1/2}$ is a finite renormalized determinant of the Dirac operator on a smooth manifold. In contrast to bosons this result is independent of the short-distace behavior of the Green's function of $\psi$ and, in fact, relies on the Pauli principle banning non-trivial occupation numbers.

\section{Discussion and conclusions}
The conformal mechanism of Sect.4 and the general mechanism of Sect.5 lead to the same result -- dynamical suppression of the topological transitions characterized by ripping the spacetime patch into two smooth disconnected pieces. This type of a topology transition happens in Euclidean quantum gravity when the latter is viewed as the derivative of the physical gravity theory in the Lorentzian spacetime. At the topological level the Euclidean theory inherits from the Lorentzian theory the arrow of time and incorporates the junction conditions that make the contribution of such topologies vanishing. The question arises whether these mechanisms are equivalent.

For boson fields both mechanisms have the form of the ultraviolet logarithmic divergence. If we regulate the coincidence limit of the Green's function by a point separation method,
    \begin{eqnarray}
     G(x,y)\sim\frac1{\sigma(x,y)}\equiv
     \frac2{\Delta^2}\to\infty,\,\,\,\,
     y\to x,                            \label{pointsep}
    \end{eqnarray}
where $\sigma(x,y)=\Delta^2/2\to 0$ is the world function -- one half of the square of the geodetic distance between $x$ and $y$ -- then the suppression factor of (\ref{suppression}) reads as
    \begin{eqnarray}
    \Big(G(x_+,x_+)
    +G(x_-,x_-)-2G(x_+,x_-)\Big)^{-1/2}\sim
    \exp\big(\ln\Delta\big).
    \end{eqnarray}
In fact this logarithmic behavior can be generalized from the 4-dimensional spacetime to any dimension $D>2$ and breaks only in two dimensions when the Green's function has a logarithmic short-distance asymptotics, $G(x,y)\sim \ln\sigma(x,y)$.

The conformal anomaly suppression factor (\ref{onshell1}) of Sect.3 can be regulated by shifting the lower integration limit from $a=0$ to $\epsilon\to 0$, so that it also takes the form of the logarithmic divergence
    \begin{equation}
    \exp(-\varGamma_0)\,\Big|_{\; HH}\sim
    \exp\left(-\frac{8k m_P^2 B}3
    \int_\epsilon^{a_{\rm max}}
    \frac{da}a\right)
    \sim\exp\big(2k\beta
    \ln\epsilon\big).             \label{confregfactor}
    \end{equation}
We would not, however, speculate on the possible relation $\Delta=\epsilon^{2k\beta}$ between the UV cutoff of the theory $\Delta$ and the regulator $\epsilon$ of the ``foamy" structure of the spacetime topology, because these two divergences originate from different orders of short distance behavior. The Green's function asymptotics (\ref{pointsep}) is responsible for the quadratic UV divergence of the quantum effective action, whereas the conformal anomaly is associated with its logarithmic divergence and is controlled by the relevant coefficient (\ref{B}) in (\ref{anomaly}).

Moreover, in view of (\ref{Pauliprinciple}) fermions give identically vanishing factor that can hardly be regulated, whereas in the conformal mechanism their contribution to (\ref{confregfactor}) is controlled by $\beta$. This coefficient of the topological Gauss-Bonnet term in (\ref{anomaly}) reads
    \begin{eqnarray}
    \beta=\frac1{360}\,\big(2 N_0+11 N_{1/2}+
    124 N_{1}\big),                \label{100}
    \end{eqnarray}
where $N_0$, $N_{1/2}$ and $N_{1}$ are respectively the numbers of spin zero scalars, spin 1/2 Weyl spinors and spin 1 vector fields \cite{Duffanomaly}. Despite statistics the effect of fermions on $\beta$ has the same sign as that of bosons, so that all particles participate on equal footing in the conformal mechanism of topology change suppression. This is another indication to the discrepancy between the mechanisms of this suppression. The source of this discrepancy remains unclear and deserves further study.

There are several other issues for future research. To begin with, our mechanism works only in the one-loop approximation, and the extension to multi-loop orders can relax it or destroy its universality. There is a hope, however, that for fermion fields it will still hold due to Pauli principle. Fortunately, the prediction of the CFT cosmology with a large number of conformal species $N\gg 1$ belongs to semiclassical domain \cite{slih,DGP/CFT}, and the effect of higher order loops is unlikely to change the situation. Also, irrespective of the topological context, it would be interesting to consider a similar effect of junction conditions in Lorentzian spacetime. The analogue of the points $x_\pm$ of Eq. (\ref{suppression}) in the Lorentzian de Sitter spacetime are antipodal points at which the Green's function $G(x_+,x_-)$ is singular for a whole family of de Sitter invariant vacua \cite{Allen}. Therefore, cancelation of singularities is possible in the combination $G(x_+,x_+)+G(x_-,x_-) -2G(x_+,x_-)$. This makes the suppression effect depending on the vacuum choice and might lead to new selection rules for antipodal identifications of de Sitter and black hole spacetimes \cite{SanchezWhiting}.

Conceptually, it is important that for local boson fields the suppression effect is mediated by their UV behavior. Nevertheless, this effect cannot be excluded by UV renormalization, because it would require a point-like counterterm localized at the junction point of the topological transition. Introduction of such counterterms would be unnatural, because they would not correspond to any type of renormalization, and their localization can be ascribed only to special regular points emerging in the solutions of effective equations in a way described above in Sect.4. This situation is different from boundary counterterms which arise in models with branes of various codimensionalities and boundaries. These counterterms are also defined on subspaces  of a lower dimensionality, but they occur as a part of physical setting -- specification of an initial quantum state on a Cauchy surface or the brane action on timelike world sheets of the branes \cite{qeastb}.

It should be emphasized that dynamical suppression we suggest applies only to a limited class of topological transitions of Figs.\ref{Fig.3} and \ref{Fig.3a}. It does not include suppression of topologies with handles which imply changing the topology of 3-dimensional sections and creation of baby universes in the spirit of \cite{babyu1,babyu2}. From the viewpoint of Lorentzian theory, underlying the EQG formalism, such topology changes are kinematically forbidden, because our starting point -- the microcanonical density matrix (\ref{projector}) -- in our physical setting is simply unaware of the concept of time and evolution. It only knows a fixed topology of a 3-dimensional space (chosen above as $S^3$) -- the range of the continuous coordinate label ${\bf x}$ in the condensed index $\mu=(\perp, a, {\bf x})$ of the set of quantum constraints $\hat H_\mu$. This happens due to spatially closed nature of our model in which the total Hamiltonian reduces to the linear combination of constraints. As a result, the time parameter $t$ arises only as an operator ordering label which helps to resolve the noncommutative algebra of $\hat H_\mu$ in (\ref{projector}) in the form of the canonical path integral \cite{PhysRep,why}. This construction of the density matrix (\ref{rhocanonical}) at the kinematical level leaves us with the only set of topological transitions of the pinching type, shown on Figs.\ref{Fig.3} and \ref{Fig.3a}. In view of continuity of integration histories this enforces the junction conditions (\ref{junction}) at the pinching points. Then, as a consequence a dynamical mechanism of topology suppression enters the game and excludes vacuum configurations from the statistical sum.

Of course, one can adopt a less conservative standpoint and regard EQG as a fundamental first principle of the theory in the spirit of \cite{noboundary}. Then path integration includes disconnected topologies without any junction conditions for quantum fields, which will restore the contribution of no-boundary instantons with finite nonvanishing one-loop prefactors. This would lead to two families of multiple sphere de Sitter instantons $\bigcup_{i=1}^k S_i^4$ with the Hubble factors (\ref{Hpm}) and all known difficulties associated with their negative on-shell action $\sim -k/H_\pm^2$. This is the infrared catastrophe of $H_-\to 0$ and divergent contribution of $k\to\infty$ configurations weighted by $\exp(\# km_P^2/H_\pm^2)$ \cite{mult-inst}. They might perhaps be cured within the tunneling prescription for the cosmological wavefunction by inverting the sign of the action \cite{tunnel}. Though this prescription was recently endowed with the EQG path integral formulation and, moreover, has interesting application in Higgs inflation \cite{tunnelEQG}, its status remains somewhat questionable.\footnote{Tunneling prescription within the density matrix approach is characterized by the deformation of the integration contour (\ref{imaginaryaxis}) to the left half of the complex plane of the Euclidean lapse $N$ -- opposite to the case of saddle points with a real $N>0$ considered above. Effectively this leads to flipping the sign of the action only for heavy massive quantum fields treated within gradient expansion. This effect does not work for massless conformal fields and also leaves unanswered certain subtle questions of analytic continuation \cite{tunnelEQG}.}

Altogether, this does not strengthen the status of EQG as a first principle concept and gives strong priority to topology suppression mechanisms rooted in Lorentzian theory. This leaves us with the thermal quantum state for the primordial power spectrum of cosmological perturbations.  Thus, it opens a new thermal mechanism for the red tilt of the CMB anisotropy, complementary to its widely accepted origin from the vacuum state \cite{ChibisovMukhanov}.

\section*{Acknowledgements}
I am grateful to D. Gorbunov, D. Levkov and especially to V. A. Rubakov for thought provoking discussions and deeply appreciate critical questions and remarks by J. B. Hartle. This work was supported by the RFBR grant No. 11-02-00512.

\end{document}